# Graphene Nanoribbon based T Junctions

Fangping   OuYang

*School of Physics Science and Technology, Central South University, Changsha 410083, P. R. China*

Bing Huang,   Zuanyi Li,

*Department of Physics, Tsinghua University, Beijing 100084, P. R. China*

Xiao jin,   Hui Xu[†]

*School of Physics Science and Technology, Central South University, Changsha 410083, P. R. China*

( Dated: January 16, 2008 )

Graphene nanoribbons (GNRs) based T junctions were designed and studied in this paper. These junctions were made up of shoulders (zigzag GNRs) joined with stems (armchair GNRs). We demonstrated the intrinsic transport properties and effective boron (or nitrogen) doping of the junctions by using first principles quantum transport simulation. Several interesting results were found: i) The I-V characteristics of the pure-carbon T junctions were shown to obey Ohm's law and the electrical conductivity of the junction depends on the height of the stem sensitively. ii) boron (or nitrogen) doping on the stems doesn't change the Ohm's law of the T junctions, but the result is opposite when doping process occurs at the shoulders. This feature could make such quasi-2D T junction a possible candidate for nanoscale junction devices in a 2D network of nanoelectronic devices in which conducting pathways can be controlled.

---

[†]Author to whom correspondence should be addressed.
Email address: oyfp04@mails.tsinghua.edu.cn

Despite its short history, graphene [1] is considered as a promising material for electronics by both academic and industrial researchers.[2-5] Several experimental groups have successfully produced isolated, and stable at room temperature, 2D crystals of graphene and graphene nanoribbons (GNRs).[6-11] Due to its novel electronic properties[2-3,12] and potentials for future applications in nanoelectronics and spintronics, graphene has attracted extensive attention. Recently, as the first step toward more complex graphene-based devices, a number of GNR-based nanoelectronic devices have been fabricated and studied theoretically, such as GNR-FETs[13-14], *p-n* junctions[15-18], nanosensors[19-20] and spin filter devices[21]. This fact further improves the prospects of graphene-based electronics.

Applied as universal joints, T-type junctions [22-23] would be possible building blocks for nanoscale devices forming a 2D network. Unlike the simple tube bends, the T-type junction is in reality a quasi-2D junction. Very recently, since GNRs as narrow as 20~10 nm[24-25] wide have been fabricated by e-beam lithography and etching techniques, it is more interesting and practical to obtain more comprehensive information about the physical and chemical properties of T-type junctions base upon GNRs and to predict possible applications

Herein, we propose a new type of T-type junction made up different kinds of GNRs by using first-principles quantum transport calculations. Several interesting computational results have been obtained: i) The I-V characteristics of the pure-carbon T junctions were shown to obey Ohm's law and the electrical conductivity of the junction depends on the height of the stem sensitively. ii) boron (or nitrogen) doping on the stems don't change the Ohm's law of the T junctions, but the case is different when doping process occurs at the shoulders. Our results provide a new method to build graphene based electronics devices.

We carry out first-principles transport calculations in our work using an *ab initio* code package, Atomistix ToolKit 2.0 (ATK),[26-27] which is based on real-space, nonequilibrium Green's function (NEGF) formalism and the density-functional theory (DFT). The calculation of the complete system can be obtained from two independent calculations of both electrode regions, and a two probe calculation of the central scattering region. The electrode calculations are performed under periodical boundary conditions, with the unit cell being six atomic layers along the ribbon axis and the *k*-point grid being 1x1x50. Self-consistent calculations are performed with a mixing rate set to be 0.02, and the mesh cutoff of carbon atom is 120 Ry to achieve a balance between calculation efficiency and accuracy. The approximation for the exchange-correlation functional is the spin-unpolarized generalized gradient approximation (GGA). The C-C and C-H bond lengths are set to be 1.42 Å and 1.1Å, and these values had been test in our previous work.[14]

Note that the definition of GNRs in our work is accordance with previous convention [28-29]: The zigzag (or armchair) GNRs are classed by the number of C-C chains forming the width of the ribbon, and the ZGNR (AGNR) with *n* C-C chains is named as *n*-ZGNR (*n*-AGNR).

It's well known that all the ZGNRs are metal at finite temperature, while all the AGNRs are semiconductor, so all ZGNRs can be used as metal leads when we build graphene-based electronic devices. What's more important, our previous results show that all the ZGNRs and AGNRs have the same work function[14], which indicate that the contact resistance can be avoided when we design the junction using different kinds of GNRs.

In Fig. 1 we show three kinds of T junction devices: Fig. 1a is the undoped GNR T junction, which is made up of metallic shoulder (ZGNR) and semiconducting stem (AGNR). The similar structures in CNTs have been studied widely[22-23] and these structures have been used as rectifier valves. It is thus natural to do a similar investigation for GNR based T junction, and the main aim of this article is to discuss the feasibility of using GNR T junctions as electronic devices.

Nitrogen (N) and Boron (B) atoms are typical substitutional dopants in carbon materials, and their binding with C is covalent and quite strong, comparing to that of host C-C bond. The incorporation of N or B atoms into the carbon materials will influence the electronic and transport properties of C host by introducing extra carries and new scattering centers. So it is very important to investigate the effect of B (or N) doping in the performance of T junctions. Fig. 1b show the T junction with stem doped by B (or N) and Fig. 1c is similar to Fig. 1b but with shoulder doped by impurities.

Fig. 2 shows the conductance and I-V characteristics of pure T junctions. From the I-V curve we found the pure T junction show a perfect metallic behavior. This is also consistent with the results of density of states (DOS) in Fig. 2a. there is a high peak near Fermi level which is contributed by the edge states of shoulder. Fig. 2c shows that the conductance of this system depends on the height of the stem, and the conductance decreased quickly when the height of stem increased. When we increased the height of the stem, due to the semiconducting behavior of stem, the resistance will also increase, and this will decrease the electron resistance. This phenomenon is interesting and indicates that we can control the transport property of T junction by selecting the length of stem in experiments.

Next we will study the B or N doped T junctions, as shown in Fig. 3. In Fig. 3c we show the conductance of pure T junction and doped T junctions. Compared with the pure T junction, we found that if doping process occurs at the stem of the junction, the conductance at Fermi level will just decrease slightly. But if the impurities are selectively doped at the shoulder of this junction, the conductance near Fermi level will decrease notably (from 0.9 $G_0$ to 0). The above results indicate that selective doping in T junction could be used to control the conductance of this device. In Fig. 3a and Fig. 3b we show the I-V curves of B and N doping at different part of the junction. If impurities are doped at the stem part, the I-V curves changed a little, but if we dope the impurities at the shoulder part, the I-V curves changed greatly: The current is about zero when the voltage ranges from -0.5 v~0.5v and then increases slowly. The I-V curves are consistent with our conductance analysis. When B/N is doped at the stem, the conductance of the shoulder will keep the same as before, so a little changes will happen. But when the doping site is selected at the shoulder, the dopants will scatter the electrons getting through the shoulder, and the conductance of this system will decrease remarkably. This means that we can control the conductance of this system by selective doping to realize the ON/OFF states in experiments.

In conclusion, we have proposed a new type of T junction made up only of carbon atoms. By using first-principles quantum transport calculations, we investigate the transport properties and edge doping-effect of such junction made up of metallic nanoribbon (shoulder) joined with semiconducting one (stem). The tunneling currents of pure-carbon T junctionsare shown to obey Ohm's law and the electrical conductivity of the junction sensitively depends upon the height of the stem. Moreover, doping on the semiconducting portion（doping on the edges of the stem）of T junctions doesn't change the Ohm's law of the T junctions. Either *n*-type or *p*-type doping on the metallic portion (doping on the edges of the shoulder) of the T junction should yield Schottky barrier-type devices. Furthermore, we have interpreted the conductance mechanism for such junction devices. Since the T junctions can be used as "universal joints" for forming a 2D network in which conduction pathways and doping via manipulation of edge terminations can be controlled, we expect that GNRTJs would be possible applied as building blocks for nanoscale tunnel junction devices in a 2D network of nanoelectronic devices built on patterned GNRs. Unlike the tube, the T junction is in reality a 2D junction. Currently, GNRs as small as 20nm[24] and 10-15nm[25] wide has been fabricated by e-beam lithography and etching techniques，these junctions based upon patterned GNRs could be the prototypes of nanoscale tunnel devices in near future.

**Acknowledgment.**

This work was supported by the National Natural Science Foundation of China (Grant Nos. 10325415 and 50504017) and the Science Develop Foundation of Central South University(Grant No. 08SDF02). The numerical calculation was carried out by the computer facilities at Department of Physics of Tsinghua University.

**Figure Captions**

**FIG 1.**: (Color online) Schematic view of the T-junction devices built up with metallic GNR (shoulder) and semiconducting GNR (stem): graphene is contacted by source (S) and drain (D) electrodes. (a) pure T-junction,(b) T-junction with N-doped stem,(c) T-junction with N-doped shoulder. The grey, black, and white balls correspond to carbon, Nitrogen (Boron), and hydrogen atoms, respectively.

**FIG 2.**: (Color online) Projected DOS (a) and the *I-Vsd* curves (b) of the pure-carbon T junction shown in Figure1a under gate voltage 0 eV; (c) Conductance as a function of the height of the stem shown in Figure1a.

**FIG 3.**: (Color online) The *I-Vsd* curves and transmission of the T junctions shown in Figure1.

**FIG 1.**

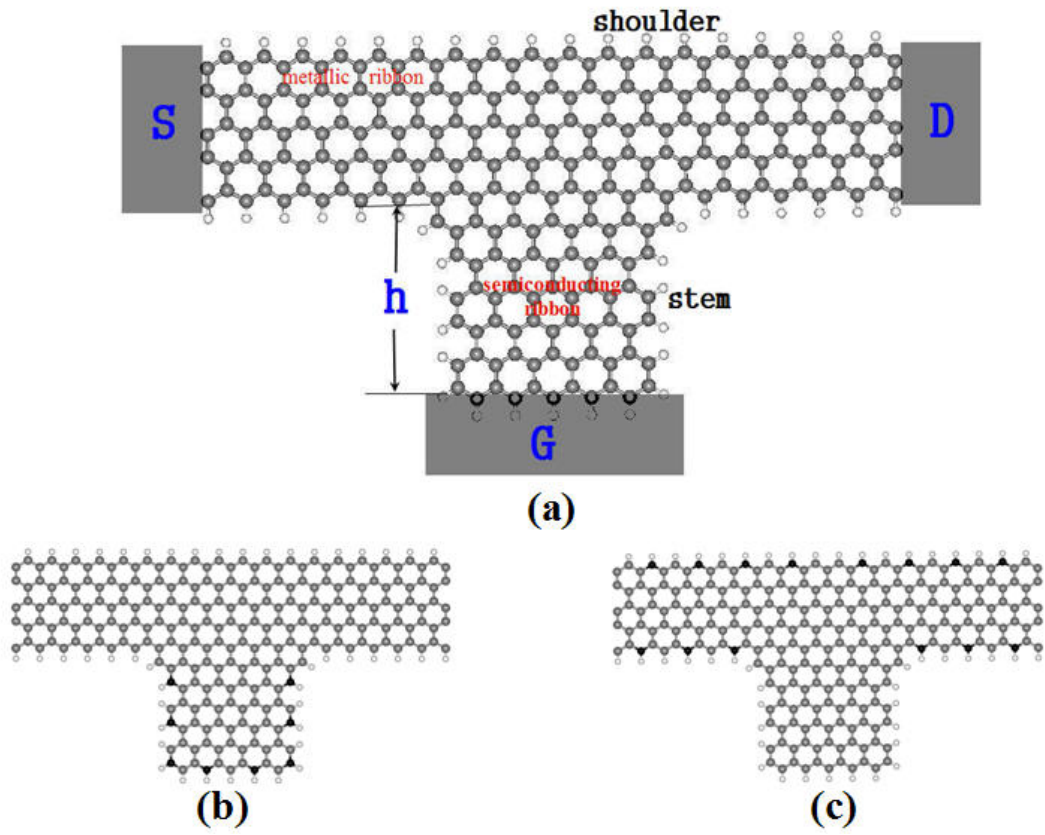

(a)

(b)          (c)

**FIG 2.**

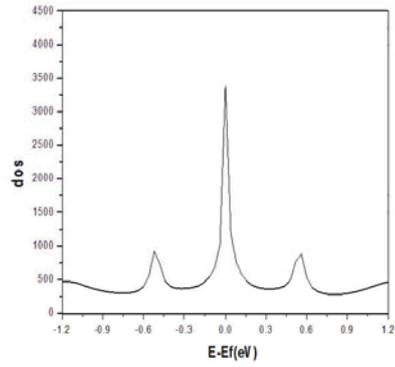

(a)

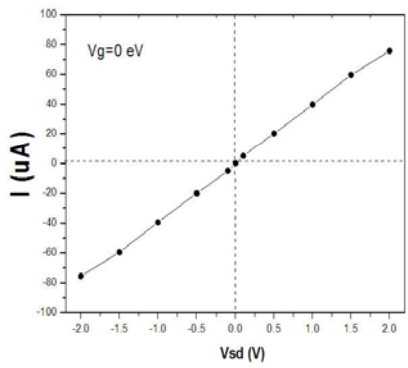

(b)

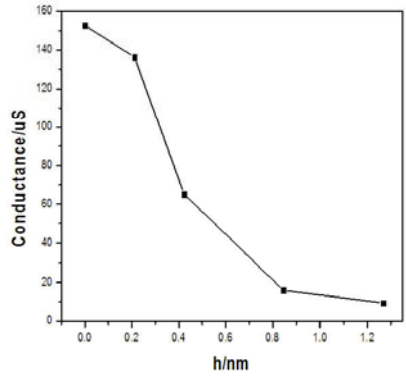

(c)

**FIG 3.**

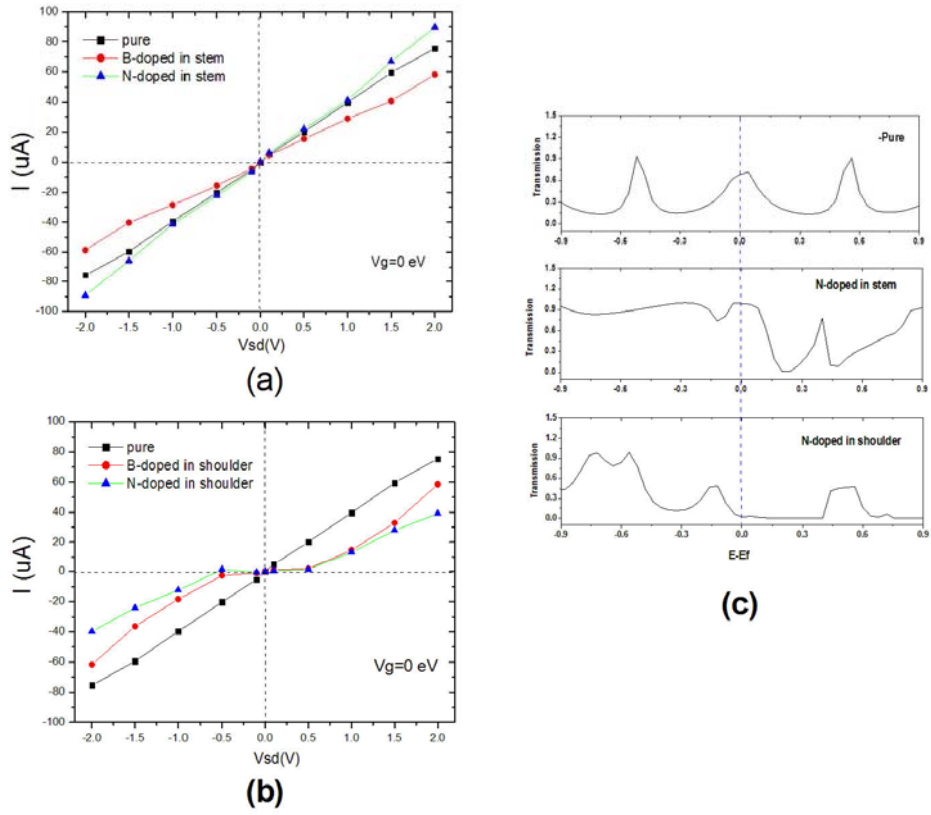